\documentclass[useAMS,usenatbib]{mn2e}

\usepackage{graphicx}
\usepackage{bm}
\usepackage{psfrag}
\usepackage{mathrsfs}
\usepackage{amssymb,amsmath}
\newcommand{\be}{\begin{eqnarray} }
\newcommand{\ee}{ \end{eqnarray} }

\newcommand{\nodata}{...}
\newcommand{\pasa}{PASA}
\newcommand{\apj}{ApJ}
\newcommand{\apjs}{ApJS}
\newcommand{\apjl}{ApJL}

\newcommand{\mnras}{MNRAS}

\newcommand{\nat}{Nature}
\newcommand{\iaucirc}{IAU circ.}

\renewcommand{\footnotesize}{\normalsize}

\title[Phase-coherent timing of the Vela pulsar]{Characterising the rotational irregularities of the Vela pulsar from  21~yr of phase-coherent timing}

\author[R.~M.~Shannon et al. ]{ R.~M.~Shannon\thanks{E-mail: ryan.shannon@csiro.au}$^{1,2}$,  L.~T. Lentati$^3$, M.~Kerr$^1$,  S.~Johnston$^1$, G.~Hobbs$^1$, \newauthor \& R.~N.~Manchester$^1$\\
$^1$ CSIRO Astronomy and Space Science, Australia Telescope National Facility, Box 76 Epping, NSW, 1710, Australia. \\
$^2$ International Centre for Radio Astronomy Research, Curtin University, Bentley, WA 6102, Australia. \\
$^3$ Astrophysics Group, Cavendish Laboratory, JJ Thomson Avenue, Cambridge CB3 0HE, UK.}

\begin{document}

\date{Accepted XX ; in original form \today}

\pagerange{1--10} \pubyear{2015}

\maketitle

\begin{abstract}

Pulsars show two classes of rotational irregularities that can be used to understand neutron-star interiors and magnetospheres: glitches and timing noise.
Here we present an  analysis of the Vela pulsar spanning nearly $21$~yr of observation and including $8$~glitches.
We identify  the relative pulse number of all of the observations between glitches, with the only  pulse-number ambiguities existing over glitch events.    
We use the  phase coherence of the timing solution to simultaneously  model the timing noise and glitches in a Bayesian framework, allowing us to select preferred models  for both. 
We find the glitches can be described using only permanent and transient changes in spin frequency, i.e., no step changes in frequency derivative.
For all of the glitches, we only need two exponentially decaying changes in spin frequency to model the transient components.
   In contrast to previous studies, we find that  the dominant transient components  decay on a common  $\approx 1300$~d time scale, and that a larger fraction ( $\gtrsim 25\%$) of  glitch amplitudes are associated with these  transient components.  
We also detect shorter-duration transient components of $\approx 25$~d, as previously observed, but are limited in sensitivity to events with shorter durations by the cadence of our observations.
 The timing noise is well described by a steep power-law process that is independent of the glitches and  subdominant to the glitch recovery. The braking index is constrained to be $< 8$  with 95\% confidence. 
This methodology can be used to robustly measure the properties of glitches and timing noise in other pulsars.

\end{abstract}

\begin{keywords}
 pulsars: general -- pulsars:~specific (PSR~B0833$-$45)  -- stars: neutron
\end{keywords}

\section{Introduction}

Pulsars are celebrated for the predictability of the arrival times of their pulses.
The power of the pulsar-timing method is realised when a phase-coherent model of the  pulse times of arrival  (TOAs) is  achieved;  that is, when there is a solution that unambiguously accounts for every rotation of the pulsar.
For example, in an analysis of $25$~yr of observations of the first millisecond pulsar, PSR~B1937$+$21, the last observed TOA in $2010$ is 518,595,058,289~periods after   first TOA in $1986$ \cite[][]{2013ApJ...766....5S}.
Phase coherence enables the detection of subtle effects that only  slightly modify the pulse frequency to accumulate in the arriving phase of pulses.  These effects include variations in the orbit of binary pulsars associated with general-relativistic effects and potentially the  passage of gravitational waves
with frequencies in the nanohertz range.
However, nearly every pulsar shows evidence for intrinsic spin irregularities which also alter the TOAs.  Irregularities are phenomenologically bifurcated into two forms, timing noise and glitches, with glitches more common in younger pulsars. 

Timing noise manifests as a red-noise (time-correlated) process in  the TOAs, and is  typically described by a wide-sense stationary stochastic process \cite[][]{1975ApJS...29..453G},  modelled as either  random walks in the pulsar spin parameters, or  with a  more general  power spectrum. 
The origin of this noise is unclear.    While some of it may be caused by variable torques associated with changes in the pulsar magnetosphere state \cite[][]{2010Sci...329..408L},  a significant fraction is likely associated with rotational irregularities interior to the star \cite[][]{2014MNRAS.437...21M}.
The properties of timing noise vary markedly across the pulsar population, with its  strength depending on the pulsar spin frequency $\nu$ and  frequency derivative $\dot{\nu}$  \cite[][]{sc2010}, and likely other factors because of  the large dispersion levels of timing noise between pulsars.
 For example, the most stable millisecond pulsar (MSP), PSR~J1909$-$3744 shows no evidence for instabilities with phase variations limited  $\lesssim 100$~ns  ($\approx 0.3\%$ of pulse phase) over  $11$~yr \cite[][]{2015Sci...349..1522S}.   
 In contrast, the  timing noise in young pulsars and magnetars can contribute many cycles of pulse phase on week to month time scales \cite[][]{2011ApJ...730...66L},  both making  it difficult to find a phase coherent solutions  in poorly sampled data and  presenting challenges to TOA-modelling algorithms. 
 While finding  the origin of timing noise is important for understanding neutron stars,   it is also necessary to account for timing noise as  part of a general timing model for the pulsar.  This is necessary to eliminate, or at least mitigate, bias in  the estimation of other parameters in the model \cite[][]{2011MNRAS.418..561C,2013MNRAS.428.1147V,2014MNRAS.437.3004L,2015arXiv150706982K}.  
 
 In addition to exhibiting timing noise,  pulsars can  experience glitch events \cite[e.g.,][]{2011MNRAS.414.1679E,2013MNRAS.429..688Y}, in which they are observed to  suddenly change  spin state, with the most significant component being increases
  in spin frequency that can exceed $1:10^5$ \cite[][]{2011ApJ...736L..31M}. 
  Glitches have been modelled with  permanent changes in $\nu$ and $\dot{\nu}$, as well as transient components, in which changes in $\nu$  are modelled to decay (typically exponentially) on  time scales
  $\tau$.  
  Occasionally multiple glitch-decay components (with different time scales) are invoked, particularly when  high cadence (daily or higher) observations \cite[e.g.,][]{2002ApJ...564L..85D}.  
   Glitch components are often identified and characterised by searching for variation of spin frequency and frequency derivative in subsets of the data \cite[][]{2015MNRAS.446..857L}. 
   This method is suboptimal because it does not utilise the phase coherence of the pulse arrival times. 
  The study of glitches is further complicated by the presence of timing noise. 
   
    There are two prevailing theoretical models for glitches.   
 In the first, glitches are associated with the transfer of angular momentum between the superfluid interior and solid crust of the neutron star \cite[][]{1975Natur.256...25A,1984ApJ...276..325A}.
 A superfluid component is  present in the core of the neutron star and a portion pervades the inner crust \cite[][]{1969Natur.224..673B}.
The quantised angular velocity vortices can pin on the nuclear lattice of the crust. 
      As the crust and the normal fluid component of the NS spin down because of electromagnetic braking, differential angular momentum is built up between the pinned vortices and the other components. 
   Eventually an external trigger or the Magnus force \cite[e.g.,][]{2009ApJ...700.1524M}  causes the vortices to unpin and transfer angular momentum to the crust, spinning the star up.  
 In the second model, glitches are associated with star-quakes in the solid, crystalline crust of the neutron star.  The quakes are the result of changes in the equilibrium configuration as the oblate star relaxes toward a spherical state as it slows down and cools \cite[][]{1969Natur.223..597R}.     
 The latter model is presently disfavoured for most pulsars because it cannot account for  the amplitude distribution and event rate observed in glitches \cite[][]{2015IJMPD..2430008H}, though it might be suitable for a few pulsars such as the Crab pulsar. 
Empirical evidence suggests that glitches can be modelled using a deterministic signal in the pulsar timing model \cite[][]{2006MNRAS.372.1549E} with only a few parameters.  
  
Theoretical models  predict that  glitch decay time scales should be constant for individual pulsars if the underlying physics driving glitch recovery is a linear process and that the transient components should have comparable magnitudes to the permanent components \cite[][]{2015IJMPD..2430008H}.   However, the transient components hitherto  measured have a small contribution relative to the permanent component and have variable decay time scales \cite[][]{2011MNRAS.414.1679E}.

Many young pulsars also show evidence for  measurable braking that is attributed to pulsar spin down.    This is a deterministic process that  primarily manifests as a  second derivative of spin frequency $\ddot{\nu}$, and is parametrized by a braking index $n$ (where $\dot{\nu} \propto \nu^n$) .  For electromagnetic braking associated with a dipolar magnetic field $n=3$;  however measured braking indices often depart from this markedly, with contributions likely arising from  angular momentum loss from particle winds and free precession.

Here we study the  timing properties of the Vela pulsar \cite[][]{1968Natur.220..340L}, a relatively  young  (characteristic age of $\tau_c \approx 11$~kyr) pulsar that shows both large levels of timing noise and glitches at quasi-regular intervals.  
The Vela pulsar is the brightest known pulsar at decimetre wavelengths, with  period-averaged flux density  of $\sim 1$~Jy.   
The first glitch in any pulsar  was identified in the Vela pulsar \cite[][]{1969Natur.222..228R,1969Natur.222..229R} and  $16$ subsequent glitches have been detected \cite[][]{2013MNRAS.429..688Y}. 
Timing noise analysis of the pulsar \cite[][]{1980ApJ...239..640C,1988ApJ...330..847C} has typically been constrained to intervals between glitches.
Glitch analyses similarly have  been rarely conducted including the effects of timing noise. 
Similarly, the observation of a very low braking index of $n=1.4 \pm 0.2$  reported for the pulsar  \cite[][]{1996Natur.381..497L} did not account for timing noise.

To analyse both the glitches and the timing noise simultaneously, we use a timing solution for the Vela pulsar that spans $21$~yr of observation and $8$ glitch events, enabling us to examine the stationarity of the timing noise and robustly estimate glitch parameters. In section \ref{sec:obs}, we present the observational data used.  In section \ref{sec:model}, we discuss the timing analysis and Bayesian methodology employed.  In section \ref{sec:results}, we compare phenomenological models for the timing noise and the glitches and select a preferred model.  In section \ref{sec:discuss}, we discuss the implications of this model. In section \ref{sec:conclusions}, we give our conclusions. 

\section{Observations}\label{sec:obs}

Our highly heterogeneous  data set comprises $1231$ TOAs obtained with the 64-metre Parkes radio telescope between $1992$~December~12 and $2014$~January~14. 
 While observations were conducted at frequencies between $0.4$ and $23$~GHz, most were made at a central frequency of $\approx 1.4$~GHz. 
Prior to $2003$ observations were made with a series of analogue-filterbank and digital-autocorrelation spectrometers; these observations are described in detail in \cite{2000MNRAS.317..843W} and \cite{2013MNRAS.429..688Y}.
   Most recently, the pulsar has been observed with digital polyphase filterbank spectrometers as part of a programme to monitor pulsars of interest to the {\it Fermi} gamma-ray observatory \cite[][]{2010PASA...27...64W}.   These observations have monthly cadence with a central observing frequency close to $1.4$~GHz and semi-annual cadence with a dual-band system capable of observing simultaneously at  central frequencies of $0.73$ and $3.0$~GHz.    

The primary data  in this analysis are TOAs, formed by correlating  observations that have been averaged in  frequency, time and (where recorded) polarisation with a template, using the commonly applied Fourier phase gradient method, described in \cite{1992RSPTA.341..117T}, and implemented in the pulsar analysis code {\sc psrchive} \cite[][]{2004PASA...21..302H}.   
Templates were produced individually for each backend/observing-band combination using an analytic model fitted to the average profile from that combination.  
Offsets between the backends were included in the timing model, as discussed in Section \ref{sec:model}.
The cross-correlation method assumes that the data can be described by the template and additive white noise.
  For our observations, this is not the case.  
Distortions of the pulse profile, especially prevalent in older observations, are introduced both by the high flux density  (in excess of system equivalent flux density) of the pulsar, and the  large dispersion sweep of the pulsar relative to the  pulse phase and frequency resolution of the observations.

Saturation of the amplifiers, other non-linear effects in the receiver and downconversion chain, and low-bit digitisation can lead to artefacts in the pulse profile, such as  apparent negative flux density on the leading and trailing edges of the pulse \cite[][]{1998ApJ...498..365J}. 
Older observations were recorded  with analogue-filterbank spectrometers with single-bit digitisers and were especially susceptible to these artefacts. 
Additionally, the pulse profile can be artificially broadened if the dispersive delay across an individual channel bandwidth is larger than the pulse-phase resolution of the observation. 
Given the relatively narrow pulse ($2.1$~ms),  and  relatively high dispersion measure ($68$\,pc\,cm$^{-3}$), older observations conducted with wide channels at low frequency show this type of broadening.
Even in more recent observations where instrumental effects are minimised, stochasticity in the pulse shape   introduces additional timing error  \cite[referred to as pulse jitter, ][]{1985ApJS...59..343C} that limits the timing precision of the observations. 
The effects of all of these distortions are secondary to TOA variations induced by timing noise and glitch events.
 It is however necessary to account for these effects in the analysis,  in particular  when modelling transient glitch components in our sparsely sampled data set.
While we do not account for them while measuring TOAs \cite[][]{2015MNRAS.454.1058L}, we account for their effect in the pulsar timing model, as discussed in the next section.

\section{Timing Analysis}\label{sec:model}

\begin{figure}
\begin{center} 
\begin{tabular}{c}
\includegraphics[scale=0.5]{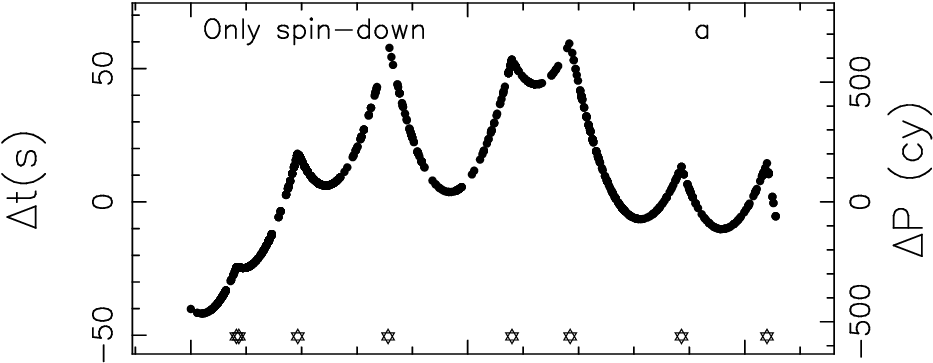}  \\
\includegraphics[scale=0.5]{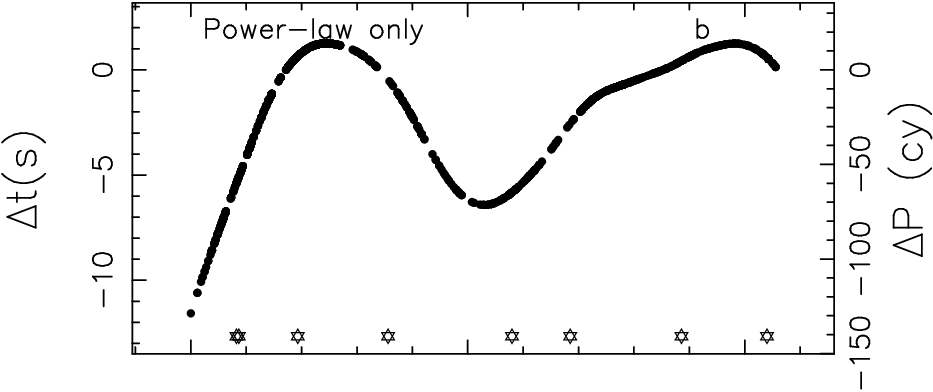} \\
\includegraphics[scale=0.5]{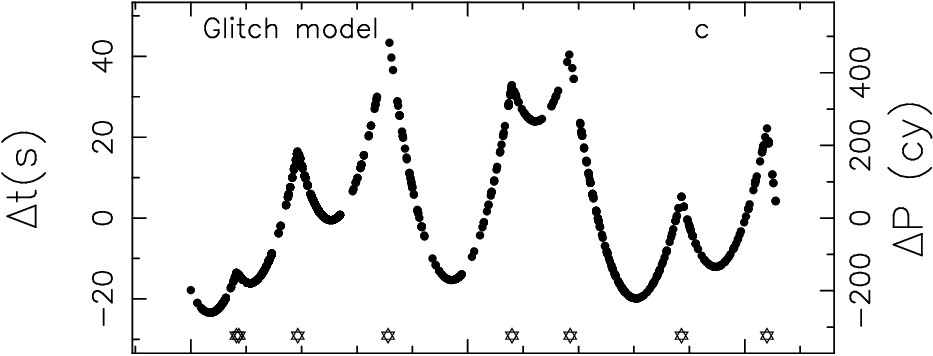}  \\
\includegraphics[scale=0.5]{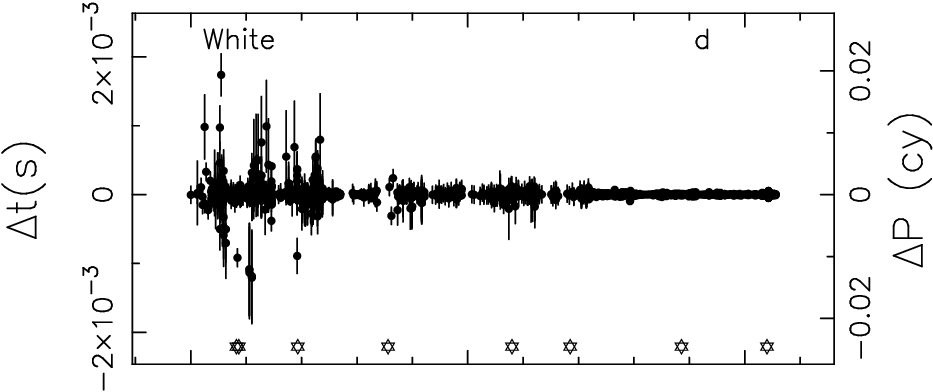}\\
\includegraphics[scale=0.5]{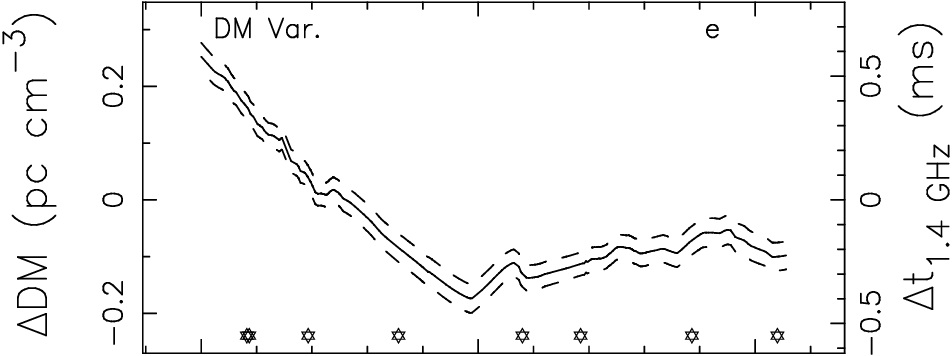} \\
\end{tabular}
\caption{  \label{fig:residual_plot}  \footnotesize Residual arrival times for maximum-likelihood models  of the Vela pulsar, measured in units of time $\Delta t$ and cycles of phase $\Delta P$.   The stars show the epochs of the glitches.   {\em a}:  Only fitting for the spin frequency and frequency derivative. 
{\em b}.    Fitting for the glitches but assuming power-law noise.   
{\em c}:  Modelled glitch signal from solution presented in panel b.    
{\em d}:  Whitened residuals for maximum-likelihood solution.
{\em e}:  Dispersion measure variations for the maximum-likelihood solution (solid line) .  The dashed line shows the $1\sigma$  uncertainties on the realisation.}  
\end{center} 
\end{figure}

The presence of strong timing noise and glitch events  make it difficult to produce phase-connected solutions  over long data spans for young, energetic pulsars like the Vela pulsar.  
As the data spans increase, the amplitude of the timing-noise signal increases rapidly \cite[with the timing noise having a power spectral density $P_r(f) \propto f^{-5 \pm 2}$  across the population, ][]{sc2010}, and relative to a spin period and period derivative at some fiducial epoch, the arrival times diverge.     Even if an initial phase-connected solution exists, it typically  has hitherto been difficult to fit the solution because small changes in model parameters  can change the residual arrival times by more than one cycle of pulse phase.



Within the pulsar-timing code {\sc tempo2}  \cite[][]{2006MNRAS.369..655H}, it is now possible to use the relative pulse numbers of the TOAs as a reference when  measuring the goodness of fit or the likelihood of a model. 
If the level of timing noise is large so that phase connecting the TOAs over the entire data set is difficult or impossible,   the relative pulse numbers for the entire data set can be determined after producing phase-connected time solutions for  subsets  of observations. 
 The overlapping intervals are chosen to be sufficiently long that the relative  pulse numbering can be checked for agreement 
 These solutions do not need to be physical, so timing noise can be whitened using sinusoids \cite[][]{2005MNRAS.360..974H} or derivatives of the pulse spin frequency.
The consistency of the solutions can be checked using TOAs common to different intervals, i.e.,  relative pulse numbers should be the same for overlapping observations.

Ambiguities in the pulse numbering potentially exist immediately following glitch events.
If the epoch of the glitch is poorly constrained, the glitch amplitude is large and the observation cadence is poor around the glitch epoch,  there may be uncertainty in the number of rotations of the pulsar between the glitch and the first post-glitch observation.  
Possible glitch epochs,  corresponding to changes in the pulse numbering by one unit, are separated by $\Delta \nu_g^{-1}$, where $\Delta \nu_g$ is the change in spin frequency associated with the glitch. 
For most of the glitches in our data set,  the glitch epochs have been previously reported  to sufficient accuracy such that we can unambiguously identify the pulse rotations  through the glitch event.
For these glitches, the maximum-likelihood glitch phase parameter is found to be within  $< 0.1$ cycles of zero  offset. 
If the glitch epoch were perfectly determined, the glitch phase parameter would be zero.  
  However, a few of the glitches epochs are poorly constrained. 
In this case we are unsure of the relative pulse numbering through the glitches.   
We tested the effects of having incorrect pulse numbers through the glitches, and found that no parameters in our analysis (beyond the glitch phase) were significantly affected, because the glitch epochs are constrained to a time much shorter than the fastest varying process in our model. 
This is to be expected, because we have included the glitch phase parameter in our timing analysis. 
Because we analytically marginalise over the glitch phase parameter, we are marginalising over the pulse-number uncertainty that exists at glitch epochs.  

Between glitches, we confirmed our pulse numbering is correct by whitening the entire dataset with a series of sinusoids \cite[][]{2005MNRAS.360..974H}.
Incorrect pulse numbering would result in TOAs that  have a random phase offset from the other arrival times, and  residuals  distributed across a full cycle of pulse phase in the whitened solution. 
We used a series of $40$ sinusoids with periods ranging from $23.2$ to $0.58$~yr.  This was sufficient to model most of the red timing noise with the residual TOAs constrained to  $\approx 0.05$ of pulse phase.  
We found no outlying points.  
In contrast, when we purposely introduced pulse numbering errors,  outlying TOAs were easily identified. 


Using this timing solution, we can directly calculate the absolute residual pulse phase of the TOAs  to the model, even if the difference between the model and data are $\gg 1$ cycle of phase.  
We used the code {\sc temponest} \cite[][]{2014MNRAS.437.3004L} to construct a complete timing model from the phase-connected solution.
The Bayesian framework implemented in {\sc temponest} enables us to simultaneously model stochastic parameters (e.g., timing noise) and deterministic parameters (e.g., glitch parameters) of interest, and marginalise over nuisance parameters of no interest to the analysis. 
For example we were able to search over a wide range of glitch decay times and glitch epochs, which is not possible with the fitting algorithm internal to {\sc tempo2}. 
 
 This  approach also enabled us  to select favoured models through the use of  Bayesian evidence (the integral of the likelihood over the parameter space weighted by the prior), which can be used to calculate the relative probabilities of  different models.
  {\sc temponest} was recently enhanced to use the {\sc PolyChord} algorithm \cite[][]{2015arXiv150201856H}  to sample posterior distributions and calculate the evidence, enabling more efficient searches and more robust calculation of evidences in high-dimensional parameter spaces than the alternative {\sc multinest} algorithm \cite[][]{2009MNRAS.398.1601F}. 

We also had to modify the code to incorporate higher precision ($128$-bit and $256$-bit) floating-point arithmetic\footnote{We implemented the arithmetic using the {\sc mlapack} library: {\tt http://mplapack.sourceforge.net/}.},   because of the  dynamic range required to model the arrival times.  The highest precision measurements have individual TOA errors (after accounting for systematics and pulse shape variations) of a few $\mu$s, while the plausible red-noise variations over the data span exceed $1000$~s.  This corresponds to a dynamic range  of $10^{18}$  for the  noise covariance.   
Matrices with this dynamic range need to be inverted as part of our analysis.  For our analysis, we found that $128$-bit precision was sufficient, and the $256$-bit precision was notably slower. 

Our model for the TOAs includes deterministic variations \cite[][]{2006MNRAS.372.1549E}  to account for the pulsar spin-down and astrometric terms.  There are  position, proper motion, and parallax measurements from long-baseline radio interferometry that exceed  the precision we can obtain through pulsar timing by factors of $\gg 10$ \cite[][]{2003ApJ...596.1137D}  and are consistent with our measurements. 
We therefore held the value fixed at the interferometrically determined position  in our analysis\footnote{While formally the VLBI astrometry should be incorporated into the prior in our analysis,  it is acceptable to fix the position at the VLBI position because it is of much greater precision, i.e., the Gaussian and delta-function priors are equivalent relative to the precision that can be measured from the data.}.
We analytically marginalised over deterministic parameters that are linear (or linearisable) in the timing model \cite[][]{2009MNRAS.395.1005V}, greatly reducing the time required to estimate non-linear and stochastic components of the timing model. 

We  also included terms to account for  stochastic time-independent (white-noise) and time-correlated (red-noise) contributions to TOAs. 
In addition to the white noise associated with the formal TOA uncertainty, terms are included to account for intrinsic shape variations and instrumental distortions, and other instrumental and astrophysical effects that are temporally uncorrelated between observations.  
We model the white noise by adjusting the uncertainty on individual TOAs to be
\be
\label{eqn:whitenoise}
\sigma^2 = F \sigma_r^2 +  \sigma_Q^2  
\ee
where $\sigma_r^2$ is the formal uncertainty derived from TOA fitting.
The factor $F$  (often referred to as EFAC\footnote{We have used the {\sc temponest} convention for defining EFAC and EQUAD which differs from the {\sc tempo2} definition.})  modifies $\sigma_R$ to account for instrumental distortions. 
 The term $\sigma_Q$ (often referred to as EQUAD)   accounts for additional observation-independent uncertainties.  
Both $F$ and $\sigma_Q$ are defined independently for each band-backend system.


Red-noise contributions to the TOAs include radio frequency-independent timing noise and dispersion-measure variations.  
Their contribution is often described by assuming the  amplitude of the fluctuations can be described using a power spectrum, which is suitable for wide-sense stationary processes.
The simplest model for timing noise we considered was a power-law power spectrum characterised with a spectral index $\beta$ and an amplitude $A$:
\be
\label{eqn:tnpl}
P_{r,{\rm PL}}(f) =  A  \left(\frac{f}{f_{\rm yr}}\right)^{\beta},
\ee
where  $f_{\rm yr}$ is a frequency of $1$ cycle per year.

In the second model, the power-law is modified  to include a spectral flattening
\be
\label{eqn:tnbl}
P_{r, {\rm BL}}(f) = \frac{A (f_c/f_{\rm yr})^{-\beta} }{\left[1 + (f/f_c)^{-\beta/2}\right]^2},
\ee
where $A$ again is the amplitude, and $f_c$ is a corner frequency.  With this definition, $P_r(f)$ is a power law when  $f \gg f_c$, and  when $f \ll f_c$, $P_r(f)$ is constant.  
This model is motivated by observations of non-power-law and apparently quasiperiodic timing  noise observed in many pulsars \cite[][]{2010MNRAS.402.1027H}.

We modelled the noise in the time domain using a harmonically related series of sinusoids constrained to have spectral density parametrized using the forms in Equations (\ref{eqn:tnpl}) or (\ref{eqn:tnbl}). 
It is necessary to mitigate spectral leakage of low-frequency power when modelling stochastic processes with Fourier series. 
However if the timing noise is band-limited or the power-law spectral index is relatively flat ($\beta >-6$), it is sufficient to start the series at $f = 1/T_{\rm span}$, because the inclusion of  $\nu$ and $\dot{\nu}$   act as a pre-whitening filter on the data set  \cite[][]{1984JApA....5..369B}.   
However, for young pulsars, timing noise can be very steep, e.g. PSR B1259$-$63 was measured to have $\beta \approx -9$ \cite[][]{2014MNRAS.437.3255S}.  
To mitigate spectral leakage in  this case, we included low frequencies $f < 1/T$ to model the lowest frequency timing noise at logarithmically spaced intervals, as described in  \cite{2015MNRAS.446.1170V}.

We also searched for dispersion measure (DM) variations. 
Following a technique outlined in \cite{2015arXiv150403692L},  we include two components in the model.
The first is a stochastic component, in which the DM fluctuations (distinct from the  TOA fluctuations) are modelled as a time series constrained to have a power-law power spectrum, \be
\label{eqn:dmpower}
P_{\rm DM}(f) = D \left( f/f_{\rm yr} \right)^{\gamma},
\ee
where $D$ is the amplitude of the DM variations and $\gamma$ is the spectral index of the assumed power law.  
The second component is a quadratic polynomial in DM that accounts for secular trends observed in some pulsars \cite[][]{2013MNRAS.429.2161K} but also acts to mitigate spectral leakage of the first component. 
Together they account for stochastic variations  associated with the the turbulent ionised interstellar medium,  but the polynomial also models linear trends in DM that are also observed  by \cite{2013MNRAS.429.2161K}.
In the TOAs, the fluctuations induced by the DM variations  scale by the inverse-square of the observing frequency, enabling DM variations and timing noise to be distinguished.

We also searched for pulsar braking,  a deterministic process that primarily induces low-frequency signals  in TOAs. 
The pulsar rotation rate  is expected to decelerate as the pulsar spins down because 
of the associated decrease in the magnetic torque.  
 As a pulsar slows down, the rate of deceleration decreases, resulting in a positive second derivative.    
In general the braking can be approximated by a braking index $n$ which modulates the secular spin evolution of the pulsar:
\be
\dot{\nu} = K \nu^n.
\ee
 In  standard magnetic-dipole braking,  the pulsar  magnetic field strength and magnetic-dipole inclination angle are assumed to be constant and $n=3$.    Departures from this value are interpreted as being associated with magnetic field evolution, changes in the spin-magnetic axis orientation, or the effects of a pulsar wind. 

For young pulsars this braking  index induces a measurable second derivative of the spin frequency $\ddot{\nu}_b$,
\be
\label{eqn:brake_nudot}
\ddot{\nu}_{b} = n \frac{\dot{\nu}^2}{\nu}, 
\ee
 and, potentially a third frequency derivative $\dddot{\nu}$,   
\be
\label{eqn:brake_nuddot}
\dddot{\nu}_{b} = n (2n-1) \frac{\dot{\nu}^3}{\nu^2}.
\ee

   Measured values of $n$  disagree markedly with  that predicted from magnetic-dipole braking.  
   For the youngest pulsars  ($\tau_c < 10$~kyr) , the braking index generally has a small positive value ($1 < n < 5$).  
   For other young pulsars ($10 < \tau_c < 10^6$~kyr), the braking index is positive, but  typically large ($n \gtrsim 10$),  suggesting that $n$  is not  associated with braking but another  process,  such as recovery from previous glitches  \cite[][]{1999MNRAS.306L..50J,2001MNRAS.328..855W,2010MNRAS.402.1027H}.       
   For the oldest pulsars ($\tau_c > 10^6$~kyr), the braking index is measured to have  both  positive and negative values, again suggesting it is not associated with pulsar braking but another process.  In these pulsars,  $n$ is likely being masked by  red noise.  
   In  previous studies of braking indices, no attempt was made to model the timing noise simultaneously to estimating  $n$. 
    We searched for pulsar braking by including it as a parameter in our timing model and adding in its  contribution to  $\ddot{\nu}$ and $\dddot{\nu}$ as described in Equations (\ref{eqn:brake_nudot}) and  (\ref{eqn:brake_nuddot}).



\section{Results} \label{sec:results}

In panel  {\em a} of Figure \ref{fig:residual_plot}, we show the residual arrival times, only fitting for the pulsar spin frequency and frequency derivative and therefore excluding terms that account for glitch events or the timing noise. 
In the absence of these terms, the residual arrival times $\Delta t$ show peak-to-peak variations of  $100$\,s, which is  $>1000$ cycles of pulse phase. 
At the epochs of the glitches (indicated by the stars at the bottom of the panel), the residuals show a discontinuous change in slope,  associated with sudden change in spin frequency. 
Between glitches, the residuals have positive curvature.
This process has been interpreted as a change in $\dot{\nu}$ and has been widely reported   \cite[][]{2011MNRAS.414.1679E,2013MNRAS.429..688Y}.  
Interestingly, the process apparently causes  the pulsar to return to comparable phase at each epoch, after accounting for the uncertainty in  $\nu$ and $\dot{\nu}$ through marginalisation.


\subsection{Models for glitches}

We  compare phenomenologically different scenarios for the glitches and timing noise, motivated by previously applied models and the observed residuals presented in panel {\em a} of Figure \ref{fig:residual_plot}:  

{\em  Permanent and  transient changes in $\nu$:}  As a minimal model, we assume that the glitches can be described by a permanent ($\Delta \nu_p$) and transient  ($\Delta \nu_t$)  change in the spin frequency. 
The permanent change introduces a change in the pulse phase at time $t>t_g$
\be
\Delta \phi(t) = \Delta \nu_p  (t-t_g),
\ee
where $t_g$ is the glitch epoch.
The transient component is assumed to exponentially  decay on a time scale $\tau$, so its contribution to pulse phase is
\be
\Delta \phi(t) =  \nu_d \tau \left[ 1-\exp\left( \frac{t-t_g}{\tau} \right) \right].
\ee

 We search over all possible values for the component amplitudes (both positive and negative) and decay times from  $1 < \tau < 10^4$~d.

 {\em Two decay time scales:}  We include an additional series of glitch decay times, to search for both long time-scale decays $\Delta \nu_\ell$ $\gg  80$~d and short time-scale glitch decay components $\Delta \nu_s$, which, respectively have decay time-scales $\tau_\ell$ and $\tau_s$. 

 {\em Common glitch-decay time scales:} In addition to assuming that the glitches can be described by permanent and transient changes in spin frequency, we assume that the transient components (long and short) all have the same time scales $\tau_s$ and $\tau_\ell$, respectively.  
 
 {\em   Permanent changes in $\dot{\nu}$:}  We assume that there are discrete changes in the spin frequency derivative $\Delta \dot{\nu}_p$ at the epoch of each glitch, as has been previously reported for the pulsar \cite[][]{2013MNRAS.429..688Y} and commonly modelled. 
The change in pulse phase is, for $t>t_g$,
\be
\Delta \phi (t)  = \frac{1}{2}  \Delta \dot{\nu}_p \left( t-t_g \right)^2
\ee

 {\em  Decay from glitch prior to our observations}:  We include exponential recovery associated with a glitch that occurred prior to the observation.   This would be recovery of a glitch occurring  $\approx 400$~d prior to our first observation \cite[][]{1991IAUC.5311....3F}, or to model long time-scale
 glitch recovery, which, as noted above, is hypothesised to be the source of timing noise in young pulsars.
 
{\em  Additional small glitches:}    In addition to the reported glitches, we search  the dataset for additional glitches.  This enables us to assess the apparent dichotomy between glitches and timing noise and search for micro-glitches that have been previously suggested \cite[][]{1988ApJ...330..847C}. 

In summary, the general  model for a glitch with permanent changes  $\Delta \nu_p$, and $\Delta \dot{\nu}_p$, and a single transient component $\Delta \nu_t$ that decays on a timescale $\tau$ is
\be
\Delta \phi(t) =  \Delta \Phi  +  \Delta \nu_p  (t-t_g), + \frac{1}{2}  \Delta \dot{\nu}_p \left( t-t_g \right)^2  \nonumber \\
  +\nu_d \tau \left[ 1-\exp\left( \frac{t-t_g}{\tau} \right) \right].
\ee
The phase $\Delta \Phi$ accounts for errors in  $t_g$  (if it is held fixed)  or errors in the pulse numbering at the glitch event. 

\subsection{Models for timing noise}

We consider a smaller number of models for the timing noise: 

{\em    No timing noise:}  We first consider a model containing no timing noise.  This model therefore assumes that the entirety of the time-correlated signal in the TOAs is associated with the glitches.

{\em    Power-law timing noise:}  We assume that the timing noise can be described by a wide-sense-stationary power-law process (Equation \ref{eqn:tnpl}) that is observed in many young pulsars. 

{\em   Band-limited timing noise:}  We assume that the timing noise is band limited and can be described by Equation (\ref{eqn:tnbl}), which would be the case if it was caused by state changing or a similar process. 

{\em    Non-stationary timing noise:}   We assume that in each inter-glitch period, the timing noise can be described by an independent power-law red-noise process. This model enables us to determine if the timing noise and glitch activity are correlated. 

{\em  Braking index:}  We additionally search for pulsar braking index $n$.
  
\subsection{Supported models}

For the models listed in the previous sections,
we calculated the Bayesian evidence  over the distribution of posterior parameters of interest while marginalising over the remaining nuisance parameters.

    In Table \ref{tab:evidence} we show a hierarchy of models that provided significantly improved evidence, culminating in the preferred model. 
In general, models that provide improvements  of $\Delta \log E >  3$ or have fewer parameters but comparable values of evidence are favoured.

   \begin{table*}
\begin{centering}
\caption{Model comparison}
\label{tab:evidence}
\begin{tabular}{cllrr}
\hline
\hline
  \multicolumn{1}{c}{Model} & \multicolumn{1}{c}{TN Model}  & \multicolumn{1}{c}{Glitch Model} &  \multicolumn{1}{c}{log(E)} & \multicolumn{1}{c}{ $\Delta$log(E) }   \\
\hline
1 &  No red noise & $\Delta \nu_p + \Delta \nu_t$ 	&4559.4&	-5394.0\\
2 & Red  & $\Delta \nu_p +\Delta \nu_t$	&	9953.4	&0.0	\\
3 & Red  $+$ DM 	& $\Delta \nu_p + \Delta \nu_t$	&10057.5	&104.1	\\
4  & Red $+$ DM & $\Delta \nu_p+ \Delta \dot{\nu}$ & 10018.4 & 65.0 \\ 
5 & Red $+$ DM	&(3) $+$ prior glitch 	&10099.7	&146.3	\\
6 & Red $+$ DM	& (4) $+$ $\Delta \nu_s$  + $\Delta \nu_\ell$	&10143.9	&190.5	\\
7 & Red $+$ DM	&(5)  $+$ common glitch time scales &10144.7&	191.3 \\ 
\hline
\end{tabular}
\end{centering}
\begin{flushleft}
Models for timing of the Vela pulsar and their evidence.  For each model we show the timing noise model, and the glitch models, and the total evidence, and relative evidence.   .  Timing noise model:   RN:  Red noise;  DM:  dispersion-measure variations.   Glitch models:   $\Delta \nu_p$: permanent changes in spin frequency;  $\Delta \nu_t$: transient change in spin frequency (if only one transient component included), modelled by as an exponential variation with an amplitude and a time scale $\tau$ ; $\Delta \nu_\ell$:  long duration transient component (if two component are included), with time scale $\tau_\ell$;  $\Delta \nu_s$:  short-duration transient component (if two components included) with time scale $\tau_s$; $\Delta \dot{\nu}$:  permanent changes in spin-frequency derivative. 

\end{flushleft}
\end{table*}
     
{\em 1.  Power-law  stationary timing noise.}     In panel {\em b} of Figure \ref{fig:residual_plot} we show the residuals from the maximum-likelihood model  accounting for the glitches,  modelling the timing noise to be a power-law red noise process and marginalising over the unknown $\nu$ and $\dot{\nu}$. 
The dominant signal in the residuals is red noise, which induces peak-to-peak variations in the residuals of $12$~s. 
The residuals are well described by a power-law red noise process with a spectral index of $\beta \approx -6$. 
We measured a consistent  amplitude and spectral shape for the timing noise  using the spectral-modelling algorithm presented in \cite{2011MNRAS.418..561C}, after fixing the glitch parameters at their maximum-likelihood values.  We find no evidence that the level or amplitude of timing noise varies between glitches.

{\em 2. Permanent frequency changes and two decaying component with  time-scales  (a short time scale of $\tau_s \approx 25$~d and a long time scale of $\tau_\ell \approx 1300$~d) that are common amongst the glitches.  }  The amplitude of the glitch components can be found in Table \ref{tab:glitch_params}.
For glitches $1$, $2$, $3$, and $5$, our data could not constrain the amplitudes of the short duration components $\Delta \nu_s$, likely because of poor sampling at these epochs.    
 In panel {\em c} of Figure \ref{fig:residual_plot}, we show the maximum-likelihood model of the glitch signal, after marginalising over the pulsar spin-down.     
The signal closely resembles the difference of panels {\em a} and {\em b}.   Once accounting for the timing noise (and marginalising over the uncertain $\nu$, $\dot{\nu}$,  DM, and instrumental jumps) the glitch signal appears to be relatively stationary in the residuals.     The maximum-likelihiood parameters for the glitches are displayed in Table \ref{tab:glitch_params}. 
We find no evidence for any variation in the glitch-decay time scales.
When modelled individually the long glitch decay time-scales show consistent posterior distributions. 
 These long time-scale  components have not previously been observed and have been  modelled  previously as a combination of  inter-glitch timing noise and  permanent changes in slowdown ($\Delta \dot{\nu}$).
 Models containing only short time scale glitches and permanent changes in $\dot{\nu}$ (Model 4 in Table \ref{tab:evidence}) are significantly disfavoured over models containing only transient components (Model 5). 
We also find evidence for the decay from a previous glitch (Model 5 in Table \ref{tab:evidence}). 
 This term accounts for latent exponential decay from a glitch  $400$~d prior to our observations.    
In contrast to previous analyses, we find  a significant component of each glitch event is  associated with a transient component, with  between $ 30\%$ and $80\%$ of the total change in spin frequency  associated with a transient component.

{\em 3.  No small glitches}. 
  In addition to modelling the $8$ large glitches in our data set, we searched for additional small glitches.    
We place limits on the amplitudes of other glitches of $\approx 10^{-7}$ averaged over the observing span.
Lower amplitude events are covariant with the timing noise.

{\em 4.   No permanent changes in spin frequency derivative  $\Delta \dot{\nu}$.  }
This component instead is attributed to the long time-scale glitch decay.  The positive curvature in the residuals between epochs, usually modelled as a change in $\dot{\nu}$  is better modelled by long-term recovery. 
We verified that our methods could detect differences between permanent changes in $\dot{\nu}$  and long-term decays by simulating data sets that contained either significant $\Delta \dot{\nu}_p$ or long-term decays.      The simulated data set had either the maximum-likelihood values of  $\Delta \dot{\nu}$ (from a model that did not contain long-term decays) or those from the long time scale decays (as listed in Table \ref{tab:glitch_params}) obtained from our data set.  The simulated data sets also contained the maximum-likelihood realisation of red noise from our data set and simulated TOA uncertainties corrected for EQUAD and EFAC.  In both cases the correct model was selected with $\Delta \log E >   130$ (i.e., with probabilities of  $ > 1- e^{-130}$).  

In panel {\em d} of Figure \ref{fig:residual_plot}, we show the {\em whitened} maximum-likelihood residuals, after accounting for  glitches, timing noise, and dispersion-measure variations (discussed below).  The residuals show no systematic variations in arrival times, suggesting that the model is complete, and providing further confirmation that the pulse numbering is correct.    
 The plot also shows that the data quality has greatly improved in the most recent data, due almost entirely  to instrumental improvements.


\begin{table*}
\begin{centering}
\caption{Glitch parameters}
\label{tab:glitch_params}
\begin{tabular}{clrrrrrrc}
\hline
\hline
  & \multicolumn{1}{c}{MJD}  & \multicolumn{1}{c}{$\Delta \nu_p$} &  \multicolumn{1}{c}{$\Delta \nu_\ell$} & \multicolumn{1}{c}{$\Delta \nu_s$} &  \multicolumn{1}{c}{$\Delta \nu_g/ \nu$} &  \multicolumn{1}{c}{$Q$} & \multicolumn{1}{c}{$(\Delta \nu_g/\nu)_{\rm lit}$} &  \multicolumn{1}{c}{Ref.}\\
  & & ($\mu$Hz) &  \multicolumn{1}{c}{($\mu$Hz)} &\multicolumn{1}{c}{($\mu$Hz)}& \multicolumn{1}{c}{($10^{-6}$)}&& \multicolumn{1}{c}{($10^{-6}$)} &\\
\hline
P & 48457&	\nodata	&16(4)	&\nodata	&\nodata	&\nodata	&2.715(2)	&1\\
1 & 49559&	1.8(2)	&7.8(2)	&\nodata	&0.86(3)	&0.81(4)	&0.835(2)	&2\\
2& 49591 &	1.4(3)	&0.7(2)	&\nodata	&0.19(2)	&0.3(1)	&0.199(2) &	3\\
3 &50369	&     12.3(3)	&11.5(3)	&0.11(3)	&2.14(5)	&0.48(2)	&2.11(2)	&4\\
4& 51559 &	22.7(6)	&12.1(6)	&0.18(3)	&3.12(8)	&0.35(2)	&3.152(2)	&4\\
5 &53193&	10.6(3)	&12.3(3)	&0.1(1)	&2.06(4)	&0.54(2)	&2.100	&4\\
6 &53960&	19.5(4)	&9.2(3)	&0.3(1)	&2.59(5)	&0.83(3)	&2.62	&4\\
7 &55408&	9.1(3)	&12.1(3)	&\nodata	&1.89(4)	&0.33(2)	&1.94  &5\\
8 &56555 &	20.9(2)	&13.2(2)	&0.21(2)	&3.06(4)	&0.39(1)	&3.100	&6\\
\hline
\end{tabular}
\end{centering}
\begin{flushleft}
Maximum-likelihood glitch parameters.  For each glitch, we list the MJD, the permanent change in spin frequency $\Delta \nu_p$,  the long and short glitch-decay amplitudes (respectively,  $\Delta \nu_\ell$ and $\Delta  \nu_s$) , reported in absolute value.   For glitches $1$,$2$,$7$, our data could not significantly constrain the amplitudes of the short glitch recoveries.   We also show the total change in spin frequency at the glitch epoch $\Delta \nu_g$, which is the sum of the permanent and transient components,  relative to the pulsar spin frequency $\nu$.  The values in parentheses represent the nominal $1-\sigma$ uncertainties for our measurements and, where available, previous measurements.  $Q = (\Delta \nu_\ell + \Delta \nu_s)/(\Delta \nu_s + \Delta \nu_\ell + \Delta \nu_p)$ is the fraction of the glitch that is recovered. 
   Where available we also show previously reported total glitch-decay amplitudes $\Delta \nu_{g,{\rm lit}}$. 
   The references are (1) \cite{1991IAUC.5311....3F};  (2) \cite{1994IAUC.6038....2F}; (3)  \cite{1994IAUC.6064....2F} ;  (4) \cite{2013MNRAS.429..688Y} ;  (5) \cite{2010ATel.2768....1B}; and  (6) \cite{2013ATel.5406....1B}. 
\end{flushleft}
\end{table*}


\subsection{Dispersion-measure variations}

As part of a larger study of dispersion-measure (DM) variations  young energetic pulsars, \cite{2013MNRAS.435.1610P} measured DM variations for the Vela pulsar, utilising some of the data presented here.  The DM variations  were modelled using a linearly interpolated time series described and as described in  \cite{2013MNRAS.429.2161K}.  
 We  searched for the amplitude and  power-law index associated with DM variations, as described above. 
 The results of our model for DM variations are presented in panel {\em e} of Figure \ref{fig:residual_plot} and  are consistent with those presented in  \cite{2013MNRAS.435.1610P}.
At $1.4$~GHz, the DM variations induce TOA fluctuations that are a factor of $10^4$ smaller than the timing noise.

\section{Discussion} \label{sec:discuss}

Compared to previous analyses of the glitches,  we find that a  much larger fraction of the changes in spin frequency is associated with transient decaying components. 
The dominant transient components are associated with a long time-scale recovery ($\tau \approx 1300$~d), with a time scale that is common to all glitches.   
Not surprisingly, the sum of our permanent and transient changes in spin frequency is comparable to previously published measurements of the permanent component as  displayed in Table \ref{tab:glitch_params}.
With only eight glitches in our data set, we have an insufficient sample size to determine if there is a correlation between the amplitude of the permanent and transient components to pulsar glitches. 
The two glitches observed on MJD 49559 and 49591 are significantly smaller in $\Delta \nu_p$ and $\Delta \nu_\ell$ than the other glitches but are also unusual because of their relative contemporaneity.

Many first-principle models of glitches  predict that transient components of different glitches  should decay with the same time scale
\cite[e.g.,][]{2010MNRAS.409.1253V}, because the microphysics of the neutron star, which regulates glitch decay, does not change. 


The length of the decays suggests that the recovery  (and the glitches) could be associated with the non-linear regime in the vortex-creep model for glitches. 
It has been predicted that the transition from linear to non-linear creep regime for the Vela pulsars would occur on time-scales of $\approx 1000$~d, close to our glitch time-scale of $1300$~d \cite[][]{1993ApJ...409..345A}.
While the regime may be non-linear, the glitch recovery would remain linear because the perturbation is relatively weak \cite[][]{1989ApJ...346..823A}, with $\Delta \nu_g/|\dot{\nu}| \ll \tau_\ell$.

Most of the inter-glitch TOA variations are associated with the long time-scale decay of the transient component (glitch recovery). 
This is consistent with observations of timing noise in other young pulsars  \cite[][]{2010MNRAS.402.1027H}, which show  $\ddot{\nu}>0 $, much like the Vela pulsar.
However, other pulsars show markedly different distributions of glitch waiting times.

We find that the timing noise can be described by a wide-sense stationary process. 
When we considered independent realisations of timing noise between glitches, we found that they all had consistent amplitudes and spectral indices.  
Furthermore, the evidence supported a single coherent process.
Once accounting for glitch recovery,  we exclude long time-scale  decay as being the origin of the timing noise.    The timing noise has a spectral index of $\beta = -6.0\pm0.5$~$(1\sigma)$, which is comparable to other young pulsars.  
    The stationarity of the timing noise suggests that  it is not related to the glitches. 
    Because the timing noise is subdominant and unaffected by the glitches, we have no evidence for any causal or correlated relationship between the two phenomena. 

We do not find any evidence for braking of the pulsar spin down and set an upper limit of $n<8$.  Our sensitivity to the braking index is limited by the large TOA variations induced by the glitch events, the presence of timing noise, and the covariance of braking index with these parameters. 
A previous measurement of braking index  relied only on measurements of $\nu$ and $\dot{\nu}$  at specific post-glitch epochs to measure the braking index     \cite[][]{1996Natur.381..497L}.
 Through $9$ glitches, they found measurements of $\dot{\nu}$  at a date  $\approx 150$~d after each glitch (and after apparent glitch recovery)  were consistent with a braking index of $1.4 \pm 0.2$.  We attempted to reproduce the results of \cite{1996Natur.381..497L} using our dataset, which span an independent set of glitches.
We whitened our dataset by including a series of sinusoids in our maximum-likelihood timing model (displayed in panel b of Figure \ref{fig:residual_plot}).   
We then used this model (which is the sum of the glitches, their recovery, and the timing noise) to to calculate $\dot{\nu}$ at epochs $150$~d after glitch events.
Like \cite{1996Natur.381..497L}, we find that there is a linear trend in ${\nu}$ at these epochs. 
However, we do not identify a linear trend in $\dot{\nu}$ at any epoch after the $8$~glitches in our observations. 
We derive an apparent braking index to be $0.0 \pm 0.9$ at epochs $120$~d after glitches. 
We attribute the differences in the apparent braking index to the the noise realisations in the two independent data sets.

 
Despite the high levels of rotational instability, the timing behaviour observed in Vela shares similarities to that observed in older pulsars which glitch less and have much lower levels of timing noise.  In a sample of $366$ pulsars monitored over $\gtrsim 30$~yr,   \cite{2010MNRAS.402.1027H} found a number of pulsars that showed cuspy timing events that occur quasiperiodically. These cuspy events represent increases in $\nu$,  followed by episodes of increased $\dot{\nu}$ that are interrupted (at the same residual phase) by another increase in $\nu$.  The best example of this is PSR~B1900$+$06 \cite[][]{2010MNRAS.402.1027H}.    If these cusps are interpreted as small glitches and the intervening episodes as glitch recovery, the rotational irregularities are analogous to what we observe in the Vela pulsar, but with a smaller magnitude. 

The robustness of the solution  presented here depends  on the underlying assumption that one of the model families considered is correct.
 The greatest uncertainty is  the assumption that the timing noise follows a power-law or broken power process.  
If the timing noise can be better constrained it may be possible to detect and characterise additional components or measure a significant braking index. 
These uncertainties affect both Bayesian and maximum-likelihood approaches. 
Searches for the most transient components are limited by the relatively poor (monthly) cadence of our observations.  

\section{Conclusions}\label{sec:conclusions}

We have presented a timing solution for the Vela pulsar that spans $\approx 21$~yr.    
The solution is nearly phase connected, with the only uncertainties associated with pulse numbering at glitch events. 
This solution, use of a full timing model, and Bayesian methodology has enabled the robust  parametrization  of the glitches and characterise the  spin noise. 

We have identified  dominant transient components to the glitches that decays on a common  $25$~d and $1300$~d time scales for all $8$ glitches in our dataset and  a subdominant steep red noise component.
 These methods can be applied to other young pulsars to identify long time-scale glitch decay components, characterise timing noise, and robustly measure braking indices.

\section*{Acknowledgments}

The Parkes radio telescope is part of the Australia Telescope National Facility, which is funded by the Commonwealth of Australia for operation as a National Facility managed by CSIRO.    GH is the recipient of ARC Future Fellowship FT120100595.     LTL acknowledges support from  a junior research fellowship at Trinity Hall College, Cambridge University.  RMS acknowledges travel support for this work through a John Philip early-career research award from CSIRO. 

\bibliographystyle{mn2e}

\appendix

\end{document}